\documentclass[aps,pra,twocolumn,superscriptaddress]{revtex4-2}

\usepackage{graphicx}
\usepackage{dcolumn}
\usepackage{bm}
\usepackage{amsmath}
\usepackage{amssymb}
\usepackage{latexsym}
\usepackage{epsfig}
\usepackage{amsbsy}
\usepackage{array}
\usepackage{amssymb}
\usepackage{setspace}
\usepackage{bm}
\usepackage{epstopdf}
\usepackage{subfigure}
\usepackage{subfigure}
\usepackage{color}

\def\sint{\ifmmode{- \!\!\!\!\!\! \int}
    \else{\hbox{$- \!\!\!\! \int \ $}}\fi}

\begin{document}

\title{Full dynamics of two-membrane cavity optomechanics}

\author{Wenlin Li}
\email{liwenlin@mail.neu.edu.cn}
\affiliation{College of Sciences, Northeastern University, Shenyang 110819, China}
\author{Xingli Li}
\affiliation{Department of Physics, The Chinese University of Hong Kong, Shatin, New Territories, Hong Kong, China}
\author{Yan Li}
\affiliation{Department of Physics, The Chinese University of Hong Kong, Shatin, New Territories, Hong Kong, China}
\author{Francesco Marzioni}
\affiliation{School of Science and Technology, Physics Division, University of Camerino, I-62032 Camerino (MC), Italy}
\affiliation{INFN, Sezione di Perugia,  via A. Pascoli, I-06123 Perugia, Italy}
\affiliation{Department of Physics, University of Naples “Federico II”, I-80126 Napoli, Italy}
\author{Paolo Piergentili}
\affiliation{School of Science and Technology, Physics Division, University of Camerino, I-62032 Camerino (MC), Italy}
\affiliation{INFN, Sezione di Perugia,  via A. Pascoli, I-06123 Perugia, Italy}
\author{Francesco Rasponi}
\affiliation{School of Science and Technology, Physics Division, University of Camerino, I-62032 Camerino (MC), Italy}
\affiliation{Department of Physics, University of Naples “Federico II”, I-80126 Napoli, Italy}
\author{David Vitali}
\affiliation{School of Science and Technology, Physics Division, University of Camerino, I-62032 Camerino (MC), Italy}
\affiliation{INFN, Sezione di Perugia,  via A. Pascoli, I-06123 Perugia, Italy}
\affiliation{CNR-INO, L.go Enrico Fermi 6, I-50125 Firenze, Italy}

\date{\today}
\begin{abstract}
In a two-membrane cavity optomechanical setup, two semi-transparent membranes placed within an optical Fabry-Pérot cavity yield a nontrivial dependence of the frequency of a mode of the optical cavity on the membranes' positions, which is due to interference. However, the system dynamics is typically described by a radiation pressure force treatment in which the frequency shift is expanded stopping at first order in the membrane displacements. In this paper, we study the \textit{full dynamics} of the system obtained by considering the exact nonlinear dependence of the optomechanical interaction between two membranes' vibrational modes and the driven cavity mode. We then compare this dynamics with the standard treatment based on the Hamiltonian linear interaction, and we find the conditions under which the two dynamics may significantly depart from each other. In particular, we see that a parameter regime exists in which the customary first-order treatment provides distinct and incorrect predictions for the synchronization of two self-sustained mechanical limit-cycles, and for Gaussian entanglement of the two membranes in the case of two-tone driving.
\end{abstract} 
\maketitle

\section{Introduction}
Cavity optomechanical systems (OMS), in which the light field is coupled to a mechanical mode through the radiation pressure interaction, have attracted significant attention as a physical platform over the past two decades~\cite{Aspelmeyer2014}. Various quantum phenomena have been studied and observed in OMS, including those related to fundamental quantum concepts such as Bell tests~\cite{Vivoli2016,Marinkovic2018}, weak measurement~\cite{Li2014,Carrasco2024}, parity-time symmetry~\cite{Lu2015,Xu2015,Li2017}, and quantum collapse models~\cite{Li2015}. Moreover, other phenomena such as the preparation of macroscopic/mesoscopic resonators in the ground state~\cite{Gigan2006,Arcizet2006,Wilson-Rae2007,Marquardt2007,Chan2011,Rossi2017} or in non-classical states~\cite{Vitali2007,Nunnenkamp2011,Pepper2011,Clerk2013,Tan2013,Liao2016,Ockeloen-Korppi2018,Barzanjeh2019}, generation of self-sustaining mechanical oscillations (i.e., a phonon laser)~\cite{Marquardt2006,Kemiktarak2014}, and their synchronization~\cite{Li2020,Sheng2020,Li2022} have been studied. Furthermore, non-trivial many-body phenomena such as topological phase transitions and spontaneous breaking of time translation symmetry can be observed in structured arrays composed of OMS~\cite{Ludwig2013,Qi2020,Brzezicki2025}. At the application level, OMS have unique advantages in quantum precision measurement~\cite{Tsang2010,Pikovski2011,Bariani2015,Barzanjeh2015,Schweer2022,Zhang2024}, since signals from the perturbed mechanical modes can be directly converted into quantum-limited optical signals. In the domain of quantum information processing, mechanical-optical intermodulation finds application in the preparation of optical devices capable of schemes such as photon state control~\cite{Nunnenkamp2011,Safavi-Naeini2011,Liao2013,Li2016,Li2018,Shen2021} and path allocation~\cite{Zhang2015,Chen2017}. 

The standard approach for describing the dynamics of OMS is based on an expansion at first order in the mechanical displacement of the optical mode energy~\cite{Mancini1994,Mancini1998,Law1995,Galley2013}. The corresponding first-order coefficient is just the radiation pressure force. For the simplest OMS consisting of a single-mode light field coupled to one mechanical mode, this model is sufficiently accurate as long as the single-photon optomechanical coupling and/or the optical driving is not too large. 
However, this standard first-order radiation pressure force approach may become inadequate when entering the strong coupling regime, and also in the multimode scenario, where the $m$-body ($m\geq 3$) cross-interaction among the multiple optical and mechanical modes may play a role. In fact, we will see that there exist circumstances in which the full exact dependence of the dispersive optomechanical interaction upon the position of the mechanical element must be taken into account. A first example is given by the case when OMS are employed for the high precision measurement of weak perturbation or small parameters. For instance, in the measurement of weak gravitational signals or of the tiny curvature of space-time couplings~\cite{Pikovski2011,Bawaj2015,Bonaldi2020,Yang2013}, the signal may be smaller than the neglected high-order interaction terms. The second example corresponds to the case where the mechanical mode possesses a high excitation energy, which is common in self-sustaining or chaotic dynamics~\cite{Lu2015,Li2020}, and may amplify the effect of higher-order terms in the dispersive interaction. Most attempts focused on the effect of pure quadratic optomechanical couplings~\cite{Thompson2008,Sankey2010,Karuza2013,Asjad2014,Buchman2012,Gao2015,Bullier2021,Li2023}, with the notable exception of Ref.~\cite{cattiaux2020}, which experimentally studied the effect of including higher-order corrections in the mechanical displacement in the interaction term in an electromechanical system. However, a more general, full dynamical analysis exploiting the exact dependence of the optical frequency shift upon the mechanical displacements is still missing.

In this paper, we discuss the so-called “sandwich-in-the-middle” structure~\cite{Lij2016,Piergentili2018} in which, starting from the “membrane-in-the-middle” one~\cite{Thompson2008,Buchman2012,Jayich2008}, the semi-transparent membrane etalon~\cite{Marzioni2025} divides the intracavity space into three regions, and optical interference yields a complex electric field distribution which is highly sensitive to the position of the membranes, and results in long-range collective interactions and strong optomechanical coupling~\cite{Lij2016,Xuereb2012,Xuereb2012A}. Based on these works, we study the full dynamics of the systems where the optomechanical interaction is described at all orders in the membrane displacement operators.
We compare the full dynamical model with the standard first-order optomechanical model linear in the membrane displacements over a wide range of parameter domains, focusing in particular on self-sustaining mechanical oscillations and their synchronization, and on mechanical Gaussian entanglement.

This paper is organized as follows. In Sec.~\ref{The model and the Langevin equation}, we derive the quantum Langevin equations of the system based on the full exact dispersive optomechanical interaction. In Sec.~\ref{Optomechanical coupling coefficient}, we discuss the magnitude of the discrepancy between the first-order coupling and the second-order coupling terms, and we individuate the parameter region where the standard approach no longer holds. In Sec.~\ref{Self-sustaining dynamics described by the full dynamics equations}, we quantitatively analyze the self-sustaining oscillations, mode competition, and synchronization effects of the resonators predicted by the full dynamics equation. In Sec.~\ref{Quantization of the full dynamics equations}, we study the Gaussian entanglement between the resonators induced by an appropriate two-tone driving of the cavity mode. Some remaining related discussions are given in Sec.~\ref{Discussion of other situations}, and concluding remarks are given in the last section.

\section{The model and the Langevin equation}
\label{The model and the Langevin equation}
\begin{figure}[]
\centering
\includegraphics[width=3in]{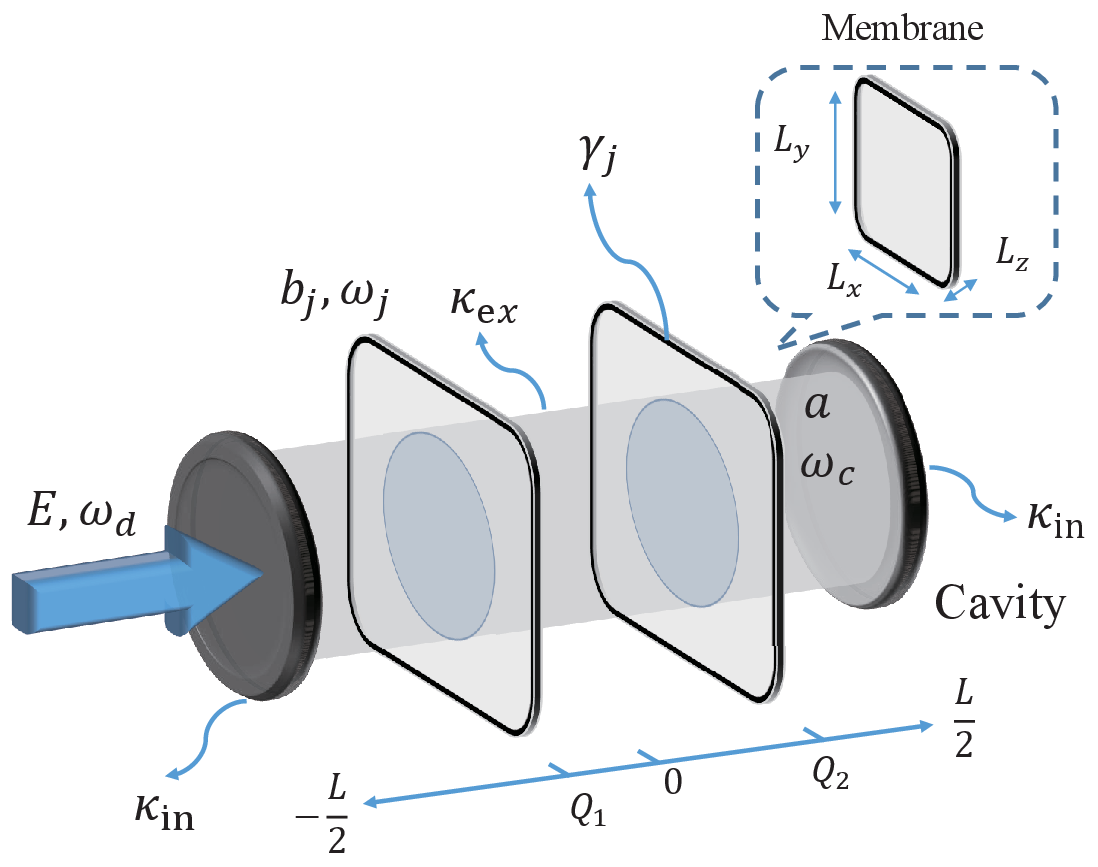}  
\caption{Schematic diagram of the system. Two movable dielectric membranes are placed inside a Fabry-P\'erot cavity of length $L$, and the driving laser field is applied from one side of the cavity. $Q_1$ and $Q_2$ are the locations where the two membranes are placed. The origin of the coordinates is set at the center of the cavity. 
\label{fig:1}}
\end{figure}
We discuss a typical two-membrane optomechanical system, comprising two movable dielectric membranes placed inside a Fabry–P\'erot cavity which is driven by an external laser. It has been demonstrated that each membrane exhibits multiple vibrational modes~\cite{Piergentili2018}. Here we focus on the case in which only one of these modes, such as the $(1,1)$ mode, has the most significant coupling with the cavity field. Then this mode is referred to as one of the ``resonators" of the OMS. The total Hamiltonian corresponding to the system is expressed as $H=H_a+\sum_jH_{mj}$, where
\begin{equation}
\begin{split}
H_a=\hbar(\omega_c+\delta \omega)\hat{a}^\dagger \hat{a}+i\hbar E(\hat{a}^\dagger e^{-i\omega_d t}-\hat{a} e^{i\omega_d t}),
\end{split}
\label{eq:freeHamilton}
\end{equation}
includes the Hamiltonian of the optical cavity and the driving field, while the free Hamiltonian of the $j$th resonator is given by:
\begin{equation}
\begin{split}
H_{mj}=\dfrac{P_j^2}{2m_j}+\dfrac{1}{2}m_j\omega_j^2X^2_j.
\end{split}
\label{eq:free_Hamilton_b}
\end{equation}
In the above expressions, $\hat{a}$ and $\hat{a}^\dagger$ are the annihilation and creation operators of the cavity mode, while $X_j$ and $P_j$ are the coordinate and momentum of each resonator relative to their respective equilibrium positions. $\omega_c$ and $\omega_j$ are the eigenfrequencies of the empty cavity and of the $j$th resonator, respectively. $E$ is related to the input laser power $P_{\text{in}}$ by $E=\sqrt{2P_{\text{in}}\kappa_{\text{in}}/\hbar\omega_c}$, where $\kappa_{\text{in}}$, corresponds to the dissipation rate of the mirror situated in the input direction. The two-membranes within the cavity (see Fig.~\ref{fig:1}) redistribute the cavity field and induce a position dependent frequency shift $\delta\omega$ of the cavity frequency, thereby resulting in the optomechanical coupling. Such frequency correction has been derived in detail in Refs.~\cite{Lij2016,Piergentili2018} from the equations for the mode resonances and we use it as starting point for our study. For the two membranes with equilibrium positions $Q_1$ and $Q_2$, the specific expression of $\delta\omega$ is
\begin{equation}
\begin{split}
&\delta\omega(Q_j,X_j)\\=&\dfrac{c}{L}\left\{(-1)^l\arcsin[\mathcal{F}(Q_j,X_j)]-\theta(Q_j,X_j)-\phi_1-\phi_2\right\},
\end{split}
\label{eq:delta_omega}
\end{equation}
where $c$ is the speed of light in vacuum and $L$ is the length of the cavity. By rewriting $X_j=q_j \sqrt{\hbar/m_j\omega_j}$ and $P_j=p_j\sqrt{\hbar m_j\omega_j}$, the function $\mathcal{F}$ and $\theta$ can be expressed in a more compact form:
\begin{equation}
\begin{split}
&\mathcal{F}(Q_j,q_j)\\=&-\dfrac{2\sqrt{R}\cos[k(\tilde{q}_1+\tilde{q}_2)]\sin[k(\tilde{q}_2-\tilde{q}_1)+\phi]}{\sqrt{1+R^2-2R\cos[2k(\tilde{q}_2-\tilde{q}_1)+2\phi]}},
\end{split}
\label{eq:F}
\end{equation}
and 
\begin{equation}
\begin{split}
&\theta(Q_j,q_j)\\=&\arcsin\left[\dfrac{R\sin[2k(\tilde{q}_2-\tilde{q}_1)+2\phi]}{\sqrt{1+R^2-2R\cos[2k(\tilde{q}_2-\tilde{q}_1)+2\phi]}}\right],
\end{split}
\label{eq:theta}
\end{equation}
where $\tilde{q}_j=Q_j+X_j=Q_j+q_j \sqrt{\hbar/m_j\omega_j}$, $k=2\pi/\lambda$ is the wave-number of the electric field and $\lambda$ is its wavelength. In these expressions we have assumed identical membranes, i.e., $R_1=R_2=R$ and $\phi_1=\phi_2=\phi$. The reflectivity amplitude of the membrane $r$ can be written in terms of the refractive index $n$ and of the thickness $L_z$ as
\begin{equation}
\begin{split}
\sqrt{R}e^{i\phi}=r=\dfrac{(n^2-1)\sin(knL_z)}{(n^2+1)\sin(knL_z)+2in\cos(knL_z)}.
\end{split}
\label{eq:R}
\end{equation}

Using the above full and exact Hamiltonian, and including fluctuation and dissipation processes for both the cavity mode and the membrane vibrational modes~\cite{Giovannetti2001}, we get the following Heisenberg-Langevin equations  
\begin{equation}
\begin{split}
&\dot{q}_j=\omega_jp_j,\\
&\dot{p}_j=-\omega_j q_j-\gamma_jp_j+\left[-\dfrac{\partial [\delta\omega]}{\partial q_j}\right]\hat{a}^\dagger \hat{a}+\xi_j,\\
&\dot{\hat{a}}=\left[-\kappa-i\omega_c-i\delta\omega\right]\hat{a}+Ee^{-i\omega_dt}+\sqrt{2\kappa}\hat{a}^{in},
\end{split}
\label{eq:quantum langevin}
\end{equation}
where $j=1,2$. Here $\gamma_j$ is the mechanical damping rate of the $j$th membrane vibrational mode, and $\kappa$ is the total decay rate of the cavity, expressed as: $\kappa=2\kappa_{\text{in}}+\kappa_{\text{ex}}$ by assuming that the two mirrors are identical, and that $\kappa_{\text{ex}}$ describes the additional field loss occurring due to scattering and absorption after the membranes are placed within the cavity. $\hat{a}^{in}$ and $\xi_j$ are the input noise terms acting on the cavity field and the $j$th membrane, respectively, whose autocorrelation functions satisfy~\cite{Giovannetti2001,noise,exp2}:
\begin{equation}
\begin{split}
&\langle \{\hat{a}^{in}(t)^\dagger,\hat{a}^{in}(t')\}\rangle= (2\bar{n}_a+1)\delta(t-t'),\\
&\langle \{\xi_j(t),\xi_j'(t')\}\rangle=2\gamma_j(2\bar{n}_{j}+1)\delta(t-t')
\end{split}
\label{eq:autocorrelation function}
\end{equation}
where $\bar{n}_{f}=[\exp(\hbar\omega_{f}/k_bT)-1]^{-1}$ is the mean thermal excitation number of the corresponding bosonic mode in thermal equilibrium with its reservoir at temperature $T$. We can explicitly write the interaction term in the equation for the mechanical modes in Eq.~\eqref{eq:quantum langevin} as
\begin{equation}
\begin{split}
-\dfrac{\partial [\delta\omega]}{\partial q_1}&=-\dfrac{\partial [\delta\omega]}{\partial \tilde{q}_1}\dfrac{\partial \tilde{q}_1}{\partial  q_1}\\&=\dfrac{c}{L}\sqrt{\dfrac{\hbar}{m_1\omega_1}}\left[(-1)^{l+1}\dfrac{\partial[ \arcsin(\mathcal F)]}{\partial \tilde{q}_1}+\dfrac{\partial[\theta]}{\partial \tilde{q}_1}\right]\\
&=\dfrac{c}{L}\sqrt{\dfrac{\hbar}{m_1\omega_1}}\left[(-1)^{l+1}\dfrac{k\sqrt{R}(f_1+f_2)}{f_3f_4}+\dfrac{kRf_5}{f_3f_6}\right]\\&:=\mathcal{L}_1(Q_1,Q_2,q_1,q_2),
\end{split}
\label{eq:L1}
\end{equation}
and 
\begin{equation}
\begin{split}
-\dfrac{\partial [\delta\omega]}{\partial q_2}&=\dfrac{c}{L}\sqrt{\dfrac{\hbar}{m_2\omega_2}}\left[(-1)^{l+1}\dfrac{k\sqrt{R}(f_7+f_8)}{f_3f_4}-\dfrac{kRf_5}{f_3f_6}\right]\\&:=\mathcal{L}_2(Q_1,Q_2,q_1,q_2).
\end{split}
\label{eq:L2}
\end{equation}
The relevant parameters in the above expressions are given in the Appendix~\ref{The renormalized dissipation rate is in the common model}~\cite{exp1}.

The vibrational motion of the membrane always satisfies the condition ${q}_{1,2}\ll Q_{1,2}$ and, expanding up to second order, one has
\begin{equation}
\begin{split}
&\delta\omega(Q_1,Q_2,q_1,q_2)\\
=&R_n+\delta\omega(Q_1,Q_2,0,0)+\sum_{j=1,2}q_j\left.\dfrac{\partial [\delta\omega]}{\partial q_j}\right|_{Q_1,Q_2,0,0}\\
&+\dfrac{1}{2}\sum_{j=1,2}q_j^2\left.\dfrac{\partial^2 [\delta\omega]}{\partial q_j^2}\right|_{Q_1,Q_2,0,0}+q_1q_2\left.\dfrac{\partial^2 [\delta\omega]}{\partial q_1\partial q_2}\right|_{Q_1,Q_2,0,0},
\end{split}
\label{eq:R1}
\end{equation}
where, except for the Lagrangian remainder $R_n$, each term has a readily discernible physical significance. The second term represents the cavity frequency shift caused simply by the static presence of the two membranes, while the third term is the standard dispersive interaction between the cavity field and the $j$th membrane, corresponding to an Hamiltonian term proportional to $\hat{a}^\dagger \hat{a} q_j$. The first two terms in the second line of Eq.~\eqref{eq:R1} describe the corresponding quadratic optomechanical coupling, which is related to a term $\hat{a}^\dagger \hat{a} q^2_j$. The last term describes a three-body interaction involving the optical cavity mode photon number and the two mechanical modes, which has rarely been considered in previous work. Substituting Eq.~\eqref{eq:R1} into Eq.~\eqref{eq:quantum langevin}, and ignoring the second order and all the other higher-order terms, we obtain the quantum Langevin equations of a standard two-membrane OMS:
\begin{equation}
\begin{split}
&\dot{q}_j=\omega_jp_j,\\
&\dot{p}_j=-\omega_j q_j-\gamma_jp_j+g_ja^\dagger a+\xi_j,\\
&\dot{\hat{a}}=\left[-\kappa-i\omega_c-i\delta\omega(Q_1,Q_2,0,0)\right]\hat{a}\\
&\,\,\,\,\,\,\,\,\,+ig_1\hat{a}q_1+ig_2\hat{a}q_2+Ee^{-i\omega_{d1}t}+\sqrt{2\kappa}\hat{a}^{in},
\end{split}
\label{eq:quantum langevin tradition}
\end{equation}
with $g_1:=\mathcal{L}_1(Q_1,Q_2,0,0)$ and $g_2:=\mathcal{L}_2(Q_1,Q_2,0,0)$ corresponding to the single-photon optomechanical coupling of a standard OMS~\cite{Aspelmeyer2014}. Moving to the frame rotating at the frequency of the driving laser $\omega_{d}$, $\hat{a}\rightarrow \hat{a} e^{-i\omega_{d}t}$, Eq.~\eqref{eq:quantum langevin} and Eq.~\eqref{eq:quantum langevin tradition} respectively become
\begin{equation}
\begin{split}
&\dot{q}_j=\omega_jp_j,\\
&\dot{p}_j=-\omega_j q_j-\gamma_jp_j+\mathcal{L}_j(Q_1,Q_2,q_1,q_2)\hat{a}^\dagger \hat{a}+\xi_j,\\
&\dot{\hat{a}}=\left\{-\kappa+i\Delta-i\left[\delta\omega\left(Q_1,Q_2,q_1,q_2\right)\right]\right\}\hat{a}+E+\sqrt{2\kappa}a^{in},
\end{split}
\label{eq:quantum langevin WA}
\end{equation}
and 
\begin{equation}
\begin{split}
&\dot{q}_j=\omega_jp_j,\\
&\dot{p}_j=-\omega_j q_j-\gamma_jp_j+g_j\hat{a}^\dagger \hat{a}+\xi_j,\\
&\dot{\hat{a}}=\left[-\kappa+i\Delta'\right]\hat{a}+ig_1\hat{a}q_1+ig_2\hat{a}q_2+E+\sqrt{2\kappa}\hat{a}^{in},
\end{split}
\label{eq:quantum langevin tradition WA}
\end{equation}
where $\Delta=\omega_d-\omega_c$ is the detuning between the drive frequency and the cavity frequency, and $\Delta'=\Delta-\delta\omega(Q_1,Q_2,0,0)$. Eq.~\eqref{eq:quantum langevin WA} and Eq.~\eqref{eq:quantum langevin tradition WA} represent the basic equations for subsequent study.
\begin{figure*}[]
\centering
\includegraphics[width=5in]{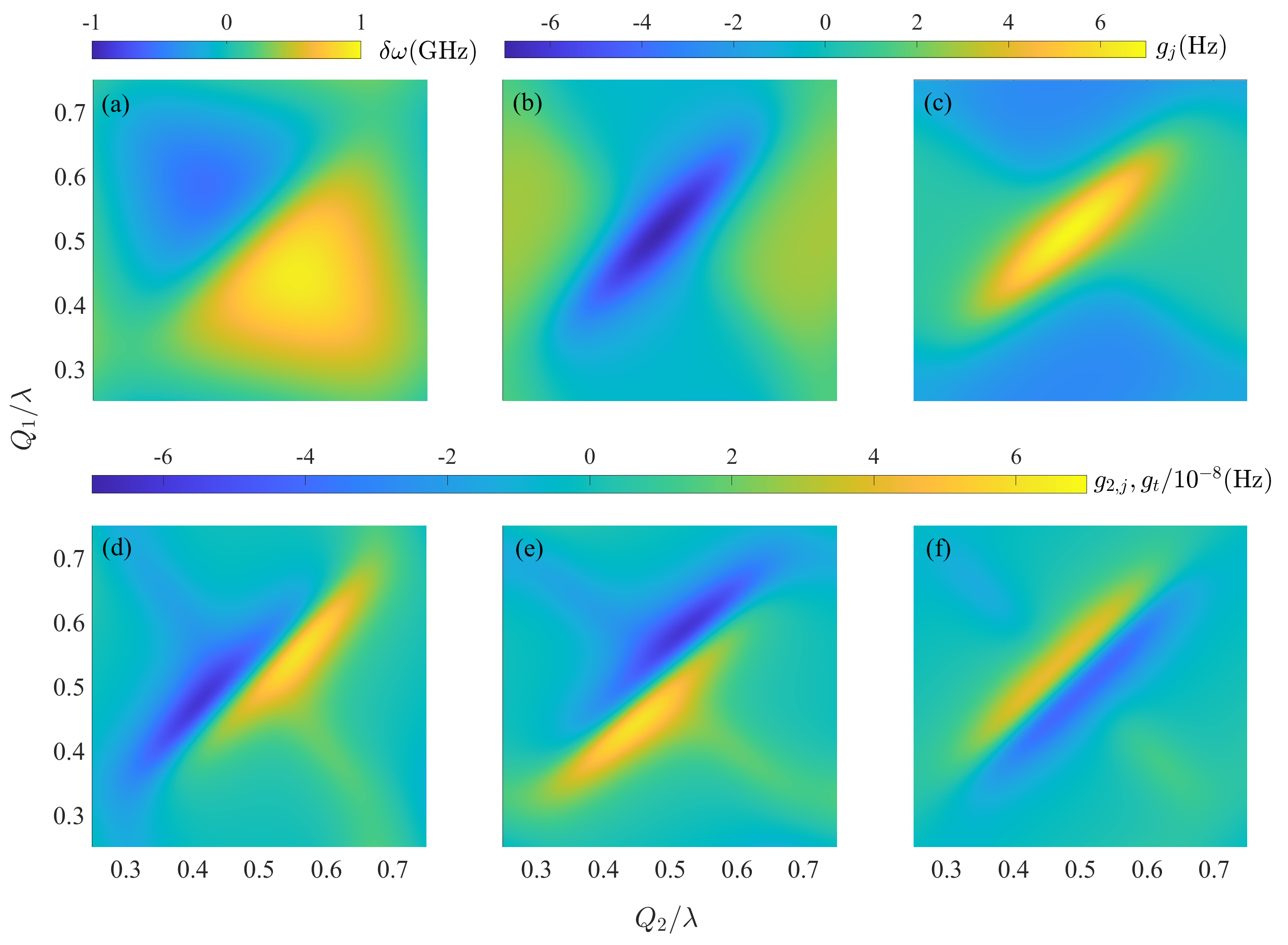}  
\caption{Coupling coefficients map in the $Q_1$-$Q_2$ plane. (a) Renormalized frequency $\delta\omega$ of the cavity field induced by the membranes. (b), (c)  First-order coupling coefficients, corresponding to the single-photon coupling strength $g_j$ of the resonator $j$. (d), (e) Quadratic coupling coefficients $g_{j,2}$ of resonator $j$. (f) Cross coupling coefficient $g_t$. Here we set $\delta=1$\,kHz and the other parameters are the same as those in Tab.~\ref{tab}. 
}
\label{fig:2}
\end{figure*}

It is important to notice that, within the framework of the standard OMS dynamics (Eqs.~\ref{eq:quantum langevin tradition WA}), the driving rate $E$, and the single-photon couplings $g_j$ are merged into a single effective parameter when analyzing semiclassical dynamics. More precisely, fixing the values of $g_1E$ and $g_2E$ implies fixing the properties of the mechanical semiclassical motion~\cite{Marquardt2006}. On the contrary, in the full exact dynamics, the driving rate $E$ is an independent parameter, and the dispersive optomechanical coupling cannot be characterized by a single parameter only. 
\section{Optomechanical coupling coefficient}
\label{Optomechanical coupling coefficient}
\begin{table}[]
\caption{Parameter}
        \centering
       \begin{tabular}{cccc}
         \hline
       Parameter & Value & Dimensionless\\ 
          \hline
         $\bar{\omega}/2\pi$  & $235.5$\,kHz & $1$&  \\
         $\vert\delta\vert/2\pi$  & $\leq 30$\,kHz & $\leq 0.1274$  \\
         $\Delta/2\pi$  & $235.5$\,kHz & $1$  \\
         $\gamma_1/2\pi$  & $1$\,Hz & $4.246\times 10^{-5}$ \\
         $\gamma_2/2\pi$  & $10$\,Hz &$4.246\times 10^{-4}$ \\
         $\kappa_{\text{in}}/2\pi$  & $50$\,kHz & $0.2123$ \\
         $\kappa_{\text{ex}}/2\pi$  & $100$\,kHz & $0.4246$ \\
         $n$& $2.17$ &  \\
         $L$& $0.09$\,m & \\      
         $\lambda$& $1064$\,nm & \\  
         $L^1_{x,y,z}$& $1.519$\,mm,$1.536$\,mm,$104$\,nm &\\
         $L^2_{x,y,z}$& $1.522$\,mm,$1.525$\,mm,$104$\,nm &\\
         $\rho$& $3100$\,kg\,m$^{-3}$  &\\
         \hline
       \end{tabular}
       \label{tab}
\end{table}
In cases where $q_j$ is not negligible in comparison to $Q_j$, Eq.~\eqref{eq:quantum langevin tradition WA} does not properly describe the dynamics of the system anymore, indeed the second order correction terms of the optomechanical coupling cannot be neglected with respect to the standard first-order ones. This will yield Langevin equations containing the quadratic optomechanical couplings:
\begin{equation}
\begin{split}
&\dot{q}_j=\omega_jp_j,\\
&\dot{p}_j=-\omega_j q_j-\gamma_jp_j+(g_j+g_{j,2}q_j+g_{t}q_{3-j})\hat{a}^\dagger \hat{a}+\xi_j,\\
&\dot{\hat{a}}=\left[-\kappa+i\Delta'\right]a+i\left(g_1+\dfrac{1}{2}g_{1,2}q_1+\dfrac{1}{2}g_{t}q_2\right)\hat{a}q_1\\
&\,\,\,\,\,\,\,\,\,\,+i\left(g_2+\dfrac{1}{2}g_{2,2}q_2+\dfrac{1}{2}g_{t}q_1\right)\hat{a}q_2+E+\sqrt{2\kappa}\hat{a}^{in},
\end{split}
\label{eq:quantum langevin tradition WA second}
\end{equation}
where $g_{j,2}:={\partial^2 [\delta\omega]}/{\partial q_j^2}$ and $g_{t}:={\partial^2 [\delta\omega]}/{\partial q_1\partial q_2}$ are the quadratic coupling strengths.  A comparison with Eq.~\eqref{eq:quantum langevin tradition WA} shows that second-order terms can be neglected only when $g_{j}/g_{j,2}\gg q_j$ and $g_{j}/g_{t}\gg q_{3-j}$ hold, and, in general, the relationship between $q_j^n$ and $g_{j,n}:={\partial^n [\delta\omega]}/{\partial q_j^n}$ determines the accuracy of the expansion. As an example, if the second-order approximation is not accurate enough, the Langevin equation obtained by Eq.~\eqref{eq:quantum langevin tradition WA second} loses any advantage over Eq.~\eqref{eq:quantum langevin WA} containing all the nonlinear dynamics. 
We are now interested in determining whether the full dynamics equations are necessary or redundant in realistic situations. We consider the specific experiment in Ref.~\cite{Piergentili2018}, whose related parameters are reported in Tab.~\ref{tab}. Here we define $\bar{\omega}=(\omega_1+\omega_2)/2$ and $\delta=\omega_2-\omega_1$. We consider $R=0.4082$, $\phi=-0.182$ and the mass of the two membranes as $m_1=L^1_xL^1_yL^1_z\rho/4\simeq188$ng and $m_2=L^2_xL^2_yL^2_z\rho/4\simeq 187$ng, corresponding to mechanical zero-point fluctuation amplitudes $\chi_{\text{ZPF}_1}=\sqrt{\hbar/m_1\omega_1}= 6.192\times 10^{-16}$\,m and $\chi_{\text{ZPF}_2}=\sqrt{\hbar/m_2\omega_2}=6.184\times 10^{-16}$\,m.  The orders of magnitude of the selected parameters are typical of membrane optomechanical systems in various experiments.

\begin{figure}[]
\centering
\includegraphics[width=3.5in]{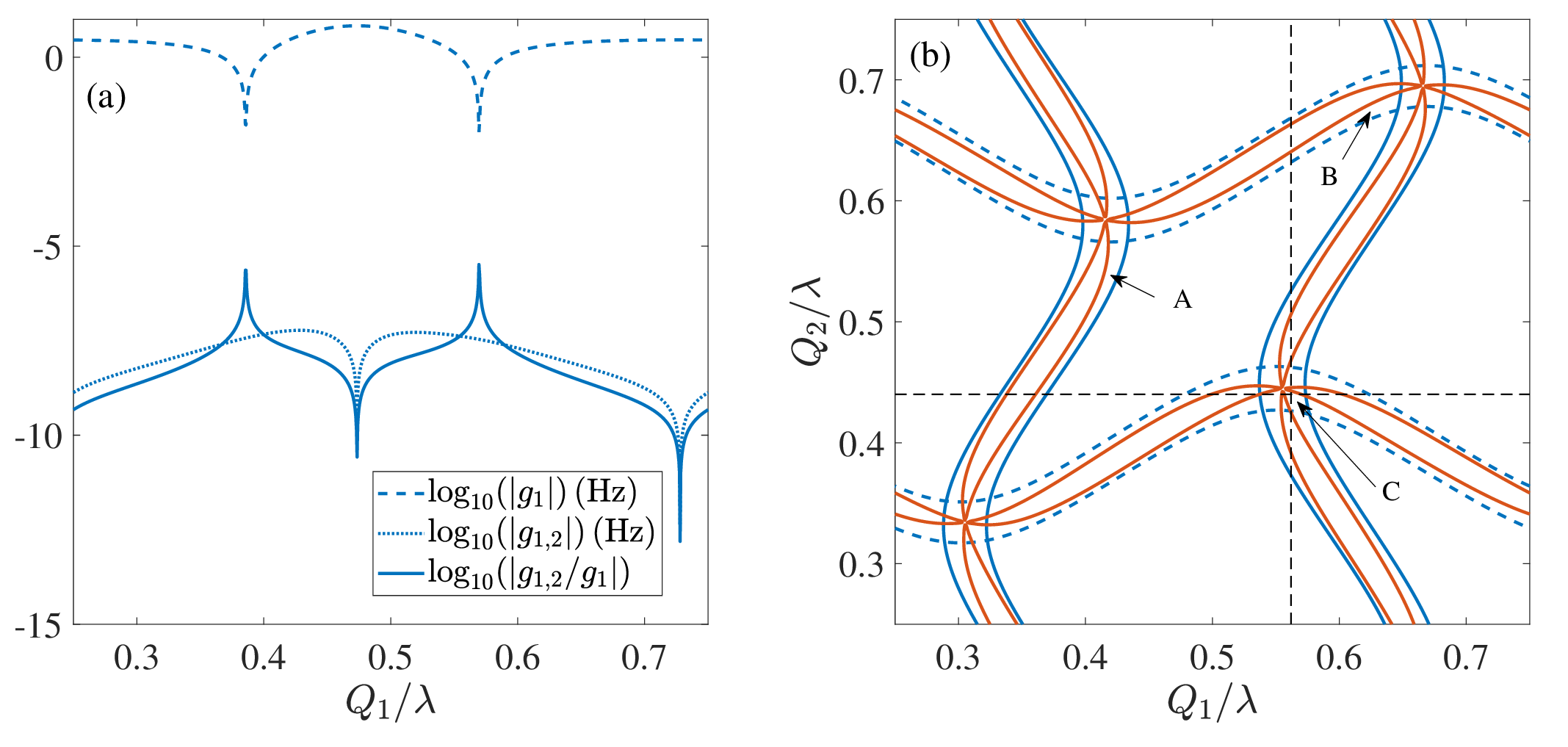}  
\caption{(a) First-order coupling strength $\vert g_1\vert$ (dashed line), second-order coupling strength $\vert g_{1,2}\vert$ (dot line), and their ratio $\vert g_{1,2}/g_1\vert$ (solid line) corresponding to oscillator $1$ as a function of $Q_1/\lambda$, with fixed $Q_2/\lambda=0.5$. (b) Contour lines of the coupling ratio on the $Q_1$-$Q_2$ plane. The blue solid lines and dashed lines represent $g_{1,2}/g_1=10^{-7.5}$ and  $g_{2,2}/g_2=10^{-7.5}$, respectively. The red lines represent both the cases for $g_{t}/g_1=10^{-7.5}$ and  $g_{t}/g_2=10^{-7.5}$. Parameters are the same as those of Fig.~\ref{fig:2}.
\label{fig:3}}
\end{figure}

In Fig.~\eqref{fig:2}, we present the coupling maps in $Q_1$-${Q}_2$ plane corresponding to $\delta \omega$, as well as the first-order coupling coefficient $g_j$ and all the second-order coupling coefficients $g_{2,j}$ and $g_t$. The coupling strengths exhibit periodicity on this basis, hence the plots correspond to a single periodic interval, $Q_j/\lambda\in[0.25,0.75]$. The coupling coefficients are significantly enhanced in those regions where the two membranes build up an effective inner cavity at resonance with the driving field. At first glance, it is evident that the first-order coupling can reach an order of magnitude of $10$\,Hz; while, the second-order coupling only up to $10^{-7}$\,Hz. Consequently, the second-order terms have been legitimately ignored in previous studies, especially when considering the oscillator in its asymptotic steady state. However, Fig.~2(b), (c) and Fig.~2(d), (e), (f) clearly show that the first and second-order couplings have different trends, which makes them closer for specific positions of the membranes. This is illustrated in Fig.~\ref{fig:3}(a), which shows the first and second-order coupling coefficients for one oscillator and their ratio as a function of position $Q_1$, while keeping fixed the position of the other membrane, $Q_2$.  The maximum value for the ratio $g_{1,2}/g_{1}$  reaches the order of magnitude of $10^{-5}$. Ref.~\cite{Li2020} investigated the self-sustaining dynamics of OMS oscillators characterized by parameters similar to those considered here, working with the standard OMS equations [Eq.~\eqref{eq:quantum langevin tradition WA}]. In this scenario, the largest displacement of the excited oscillator can readily attain a value of approximately $q_j/\chi_{\text{ZPF}_j}\sim 3\times 10^6$, which produces second (or higher)-order couplings relevant for the dynamics. It is reasonable to expect that second-order couplings cannot be ignored when the contribution they provide is higher than a correction of $10\%$ to the first-order coupling, i.e. $(g_{j,2}q_j+g_tq_{3-j})>0.1g_j$. Thus, considering $q_j/\chi_{\text{ZPF}_j}\sim 3\times 10^6=10^{6.5}$, the critical values for the coupling ratio are $g_{j,2}/g_{j}=10^{-7.5}$ and $g_t/g_{j}=10^{-7.5}$. The boundaries identified by these critical values are plotted in Fig.~\ref{fig:3}(b) as a function of the positions of the membranes. Point `A'(`B') indicates the region inside solid (dashed) blue lines where high-order coupling for oscillator $1$($2$) is not negligible, while around point `C' high-order couplings are relevant for both oscillators. Similarly, the red lines define the sector where the cross-coupling $g_t$ cannot be ignored.

Recognizing the critical regions is worthwhile when performing measurements. An hypothetical experimenter who does not consider higher-order couplings could randomly choose two positions for the membranes and consequently measure on an OMS with a blue-detuned driving. The experimenter would have approximately a $26.5\%$ probability of obtaining a discrepancy between theoretical predictions and experimental results, regardless of the accuracy of the measurements; indeed he/she would retrieve perfect correspondence between theory and experiment when applying a red-detuned driving. A detailed comparison of the effects of higher-order coupling on system dynamics will be presented in the subsequent sections of this paper.

\section{Self-sustaining dynamics described by the full dynamics equations}
\label{Self-sustaining dynamics described by the full dynamics equations}
When considering the system driven by a blue-detuned laser ($\Delta>0$), a sufficiently strong Stokes effect can break the asymptotic stability condition and excite the resonators. In this instance, the mechanical modes potentially exhibit self-sustaining or chaotic dynamics, as well as concomitant nonlinear phenomena, which can be quantitatively characterized after expressing the self-sustaining dynamics of the mechanical mode as $b_j(t):=q_j(t)+ip_j(t)=b_{0,j}+A_j(t)e^{-i\bar{\omega} t}$, where $b_{0,j}$ is a constant and $A(t)$ is a slowly-varying amplitude at the reference, average frequency $\bar{\omega}$. In this regime, both optical and mechanical amplitudes are large, the fluctuations are of thermal origin, and quantum fluctuations are negligible. Therefore, the full [Eq.~(\eqref{eq:quantum langevin WA})], and the first-order [Eq.~(\eqref{eq:quantum langevin tradition WA})] quantum Heisenberg-Langevin equations can be safely treated as standard classical Langevin equations, in which operators are replaced by ordinary functions of time.

A first non-trivial phenomenon for a single mechanical oscillator is the limit cycle which may be characterized by multiple steady states, as the driving energy increases~\cite{Marquardt2006}. However, in the presence of multiple mechanical resonators, the driving energy can be insufficient to excite all of them to the limit cycle states. In these circumstances, mode competition occurs~\cite{Kemiktarak2014,Li2020}, which may be described for two oscillators by the order parameter
\begin{equation}
\begin{split}
R_c=\log_{10}\left[\lim_{\Delta t\rightarrow\infty}\dfrac{\int_{t_s}^{t_s+\Delta t}\vert A_1(t)\vert dt}{\int_{t_s}^{t_s+\Delta t}\vert A_2(t)\vert dt}\right],
\end{split}
\label{eq:mode competition R}
\end{equation}
where $t_s$ represents the moment at which the dynamics undergo a transition from an initial state-dependent relaxation process to a steady state. Moreover, a nonlinear phase correlation may be established between the two oscillators, indicating that they have achieved spontaneous phase synchronization. This is described quantitatively by~\cite{Li2020}
\begin{equation}
\begin{split}
\lim_{t\rightarrow\infty} \dfrac{d}{dt}\cos[\theta_1(t)-\theta_2(t)]:=&\lim_{t\rightarrow\infty}\dfrac{d}{dt}\mathcal{P}(t)=0,
\end{split}
\label{eq:phase error measure}
\end{equation}
where $\theta_j(t)=\arg[A_j(t)]$ is the phase of the $j$th resonator. The degree of synchronization throughout the process can be characterized by the statistical variance of the phase difference in the time domain, that is, 
\begin{equation}
\begin{split}
\delta\mathcal{P}=\dfrac{1}{\Delta t}\int_{t_s}^{t_s+\Delta t}\left[\mathcal{P}(t)-\bar{\mathcal{P}}\right]^2dt,
\end{split}
\label{eq:phase delta}
\end{equation}
where $\bar{\mathcal{P}}$ denotes the statistical mean of the measure $\mathcal{P}(t)$, expressed as
\begin{equation}
\begin{split}
\bar{\mathcal{P}}=\dfrac{1}{\Delta t}\int_{t_s}^{t_s+\Delta t}\left[\mathcal{P}(t)\right]dt,
\end{split}
\label{eq:phase delta}
\end{equation}
and it describes a stable phase difference between the two mechanical modes when synchronization occurs. 

The discrepancy between the full dynamics equations and the standard first-order equations will be delineated in the following analysis of the aforementioned phenomena. In the considered parameter regime, all amplitudes are large, and we have verified that one can safely neglect all the noise terms in solving numerically the coupled dynamical equations.
\begin{figure}[]
\centering
\includegraphics[width=3.5in]{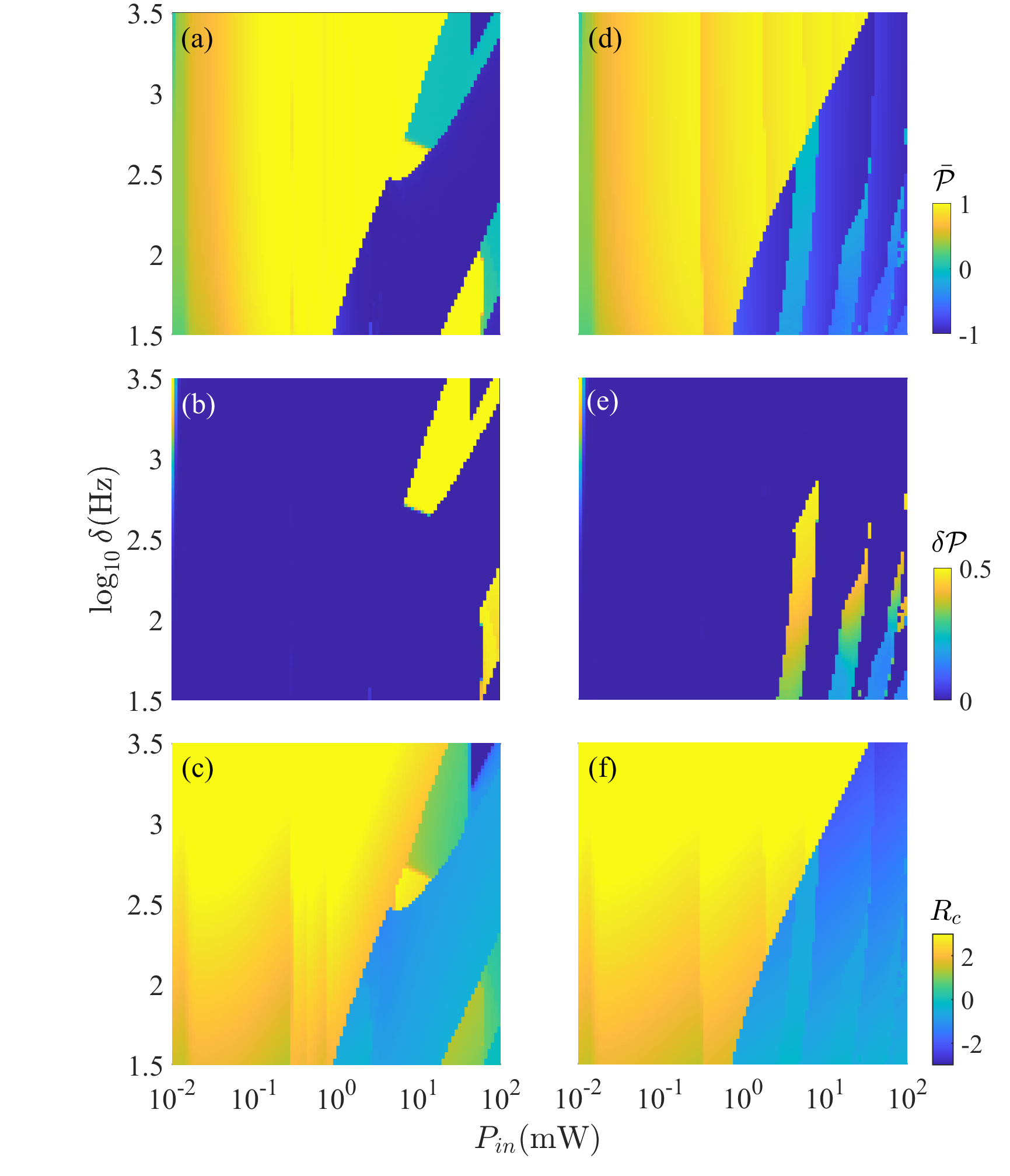}  
\caption{Synchronization phase diagrams in terms of the measure $\mathcal{P}$[(a) and (d)] and the criterion $\delta \mathcal{P}$[(b) and (e)] in the driving-detuning plane. (c), (f) The mode competition order parameter $R_c$ is represented in the driving-detuning plane. Here the phase diagrams (a), (b) and (c) are obtained from the full dynamics equations~\eqref{eq:quantum langevin WA}, while the phase diagrams (d), (e) and (f) are calculated from the first-order equations~\eqref{eq:quantum langevin tradition WA}. Here we set $\{Q_1,Q_2\}/\lambda=\{0.562,0.440\}$, $\bar{\omega}t_s= 2\times 10^6$, $\bar{\omega}\Delta t= 5\times 10^5$, the other parameters are the same as those in Fig.~\ref{fig:2}.
\label{fig:4}}
\end{figure}
\begin{figure}[]
\centering
\includegraphics[width=3in]{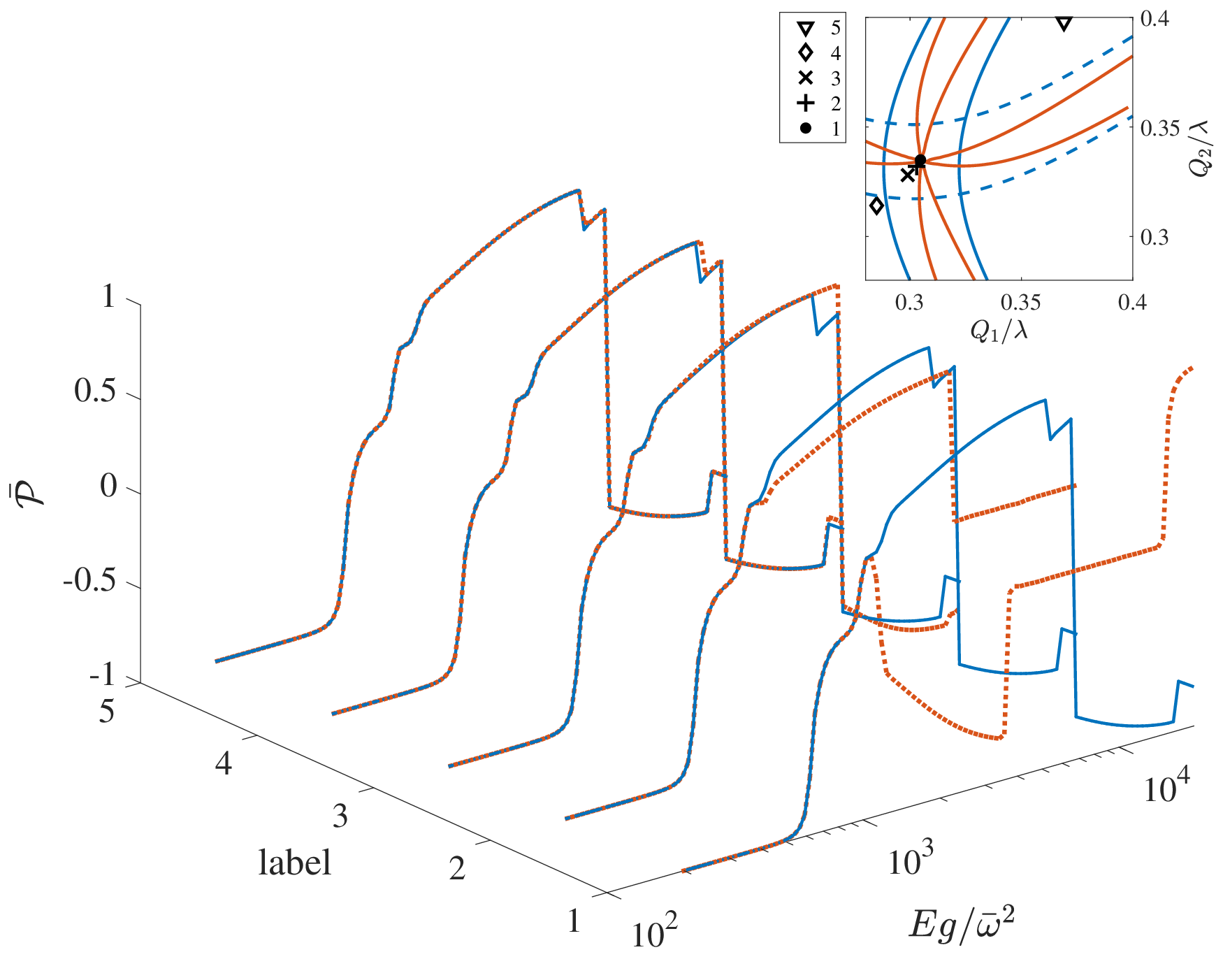}  
\caption{The synchronization measure $\bar{\mathcal{P}}$ as a function of the dimensionless parameter $Eg/\bar{\omega}^2$. Here five locations for the membranes are chosen as follows (1): $\{Q_1,Q_2\}/\lambda=\{0.3040,0.3330\}$, (2): $\{Q_1,Q_2\}/\lambda=\{0.3020,0.3310\}$, (3): $\{Q_1,Q_2\}/\lambda=\{0.2980,0.3270\}$, (4): $\{Q_1,Q_2\}/\lambda=\{0.2840,0.3131\}$ and (5): $\{Q_1,Q_2\}/\lambda=\{0.3800,0.4090\}$. In each case, the condition $g_1=g_2=g$ is satisfied. The positions in the $Q_1$-$Q_2$ plane are marked in the inset, which is a zoom of Fig.~\ref{fig:3}(b). The blue identical lines are obtained from the first-order equations~\eqref{eq:quantum langevin tradition WA}, the red lines are obtained from the full dynamics equations~\eqref{eq:quantum langevin WA}. The other parameters are the same as those in Fig.~\ref{fig:2}.
\label{fig:5}}
\end{figure}

In Fig.~\ref{fig:4}, the synchronization measure $\bar{\mathcal{P}}$, the synchronization criterion $\delta \mathcal{P}$, and the mode competition order parameter $R_c$ are plotted in the $\delta-P_{\text{in}}$ plane (resonator frequency detuning on the horizontal axis, input power on the vertical axis).  Figs.~\ref{fig:4}(a), (b) and (c) correspond to phase diagrams predicted by calculating the full dynamics equations~\eqref{eq:quantum langevin WA}, while the sub-figures (d), (e) and (f) show the results given by the first-order equations~\eqref{eq:quantum langevin tradition WA}. Here we fix the positions of the two membranes at $\{Q_1,Q_2\}/\lambda=\{0.562,0.440\}$[point `C' in Fig.~\ref{fig:3}(b)], where, according to the previous results, both the quadratic couplings $g_{j,2}$ and the cross-coupling $g_t$ become relevant. A two-by-two comparison of the phase diagrams in Fig.~\ref{fig:4} [(a)-(d), (b)-(e), (c)-(f)] reveals different trends. For weak driving [left region of the diagrams], the compared results are similar, which is consistent with the fact that the full dynamics equations regress to the first-order equations when the resonators' motion amplitudes are not large enough for the high-order terms to count. In contrast, within the range of low first-order coupling strength and high input powers, divergent trends appear, even at the qualitative level. The standard OMS equation asserts that the resonators will invariably reach a state of synchronization, because of the indirect resonators interaction mediated by the input drive. However, the full dynamics equation predicts the emergence of another transition, characterized by the loss of phase synchronization for an input power larger than a threshold value.

To further highlight the discrepancy between the two descriptions, we place the two membranes in five distinct points of the $Q_1$-$Q_2$ plane, as marked in the inset of Fig.~\ref{fig:5}. We identify them from $1$ to $5$ according to their increasing distance from the region where high-order couplings cannot be ignored. The choice of these positions guarantees that the two resonators exhibit equivalent first-order single-photon coupling strengths, i.e. $g_1=g_2=g$.  Fixing the frequency difference between the two resonators at $1$\,kHz, we calculate the synchronization measure as a function of the dimensionless parameter $Eg/\bar{\omega}^2$. The blue lines in Fig.~\ref{fig:5}, obtained from the first-order equations, provide identical dynamics for all five cases, thus supporting our discussion above. Such degeneracy of the phase synchronization is broken once the full dynamics equations are considered, as illustrated by the red lines. From point `$5$' to point `$1$', the difference between the predictions provided by the full dynamics equations and the first-order equations gradually becomes larger, until they are almost completely inconsistent for point `$1$'.

We finally observe that previous works~\cite{Kemiktarak2014,Li2020} showed that the phase transition boundary of mode competition exhibits a perfect correspondence when only first-order couplings are considered. This can be interpreted as the evolution of the phase difference $\theta_-=\theta_1-\theta_2$ satisfying a Kuramoto-type equation, whose coefficient relates the amplitude ratio $R$ of the resonators. The phenomenon is confirmed in our model by the phase diagram Figs.~\ref{fig:4}(d), (e) and (f). Furthermore, as evidenced by Figs.~\ref{fig:4}(a), (b) and (c) such correspondence remains after considering the influence of high-order couplings. We infer that the relatively weak high-order coupling strength corresponding to the selected parameters is insufficient to disrupt it. However, if extreme parameters are taken into account to induce substantial high-order couplings, the correspondence may become inconsistent.

\section{Quantization of the full dynamics equations}
\label{Quantization of the full dynamics equations}

In this Section, we compare the predictions of the full dynamics of the sandwich-in-the-middle setup [Eq.~\eqref{eq:quantum langevin WA}], with those obtained with a first-order treatment of the optomechanical interaction [Eq.~\eqref{eq:quantum langevin tradition WA}], on a very different phenomenon. In fact, here we focus on the generation of stationary entanglement between the two mechanical resonators based on a reservoir-engineering scheme~\cite{Clerk2013,Lij2015} in  which the cavity mode is bichromatically driven at two different frequencies, one close to the red sideband and the other close to the blue sideband, and described by the following Hamiltonian~\cite{Lij2015}:
\begin{equation}
\begin{split}
H=&\hbar(\omega_c+\delta \omega)\hat{a}^\dagger \hat{a}\\
&+i\hbar\left[ E_1\hat{a}^\dagger e^{-i(\omega_d+\omega_1) t}+E_2\hat{a}^\dagger e^{-i(\omega_d-\omega_2) t}\right]+h.c..
\end{split}
\label{eq:freeHamilton}
\end{equation}
This bichromatic driving tends to ground-state cool an effective mechanical Bogoljubov mode, and this is equivalent to generate a stationary Gaussian entangled state of the membrane modes (see also Ref.~\cite{Yazdi2024} for a generalized version of this reservoir engineering scheme for the generation of arbitrary Gaussian multimode entangled states of mechanical resonators).

The numerical study of the dynamics described by the Heisenberg-Langevin equations of Eq.~\eqref{eq:quantum langevin WA} and Eq.~\eqref{eq:quantum langevin tradition WA} is highly nontrivial in the general case where both the nonlinear radiation pressure interaction, and the quantum fluctuations associated with the noise terms, play a role. In fact, 
nonlinear quantum effects may be responsible for negative values of the Wigner function~\cite{Lee2013,noise}, which cannot be reproduced by the stochastic trajectories obtained from the numerical solution of Langevin equations. The most straightforward option in this case is to numerically solve the master equation for the density operator of the whole system, corresponding to the considered set of Heisenberg-Langevin equations~\cite{noise}, which is typically hard due to the large Hilbert space dimension. However, here we are restricting ourselves within a parameter regime in which a \emph{Gaussian} entangled state of the two mechanical resonators is generated, which is always described by a positive Wigner function in the phase space of the system.
Under these conditions, the quantum behavior can be safely described by means of $c$-number stochastic Langevin equations. To be more specific, we simulate the Langevin equations $N$ times, and the distribution function in phase space is then reconstructed through a statistical analysis of the results, which is equivalent to the Wigner function in this case~\cite{noise}. More precisely, the results of the $k$th simulation are recorded as $^{k}q_j$ and $^{k}p_j$, and the phase-space probability distribution function $W(q_1,p_1,q_2,p_2)$ can be evaluated as follows~\cite{Lee2013,Li2020,Wang2014}:
\begin{equation}
\begin{split}
W(q_1,p_1,q_2,p_2)=\lim_{h\rightarrow 0}\dfrac{N_{q_j,p_j}}{N h^2},
\end{split}
\label{eq:phase distribution function}
\end{equation}
where $N_{q_j,p_j}$ is the number of results satisfying $^{k}q_j\in(q_j-h/2,q_j+h/2]$ and $^{k}p_j\in(p_j-h/2,p_j+h/2]$.

A quantum Gaussian state is completely characterized by the expectation values and the covariance matrix corresponding to the complete dynamics of the system, which we express in terms of the vector of optical and mechanical observables $\hat{u}=(\hat{x}, \hat{y}, \hat{q}_1, \hat{p}_1, \hat{q}_2, \hat{p}_2)^{\top}$ with $\hat{x}=(a^\dagger+a)/\sqrt{2}$ and $\hat{y}=i(a^\dagger-a)/\sqrt{2}$. The covariance matrix is then defined as
\begin{equation}
\begin{split}
C_{i,j}=\dfrac{\langle \hat{u}_i\hat{u}_j+\hat{u}_j\hat{u}_i\rangle}{2}-\langle \hat{u}_i\rangle\langle \hat{u}_j\rangle.
\end{split}
\label{eq:c max}
\end{equation}
The covariance matrix allows us to quantify the degree of entanglement between the resonators through the \textit{logarithmic negativity} $E_n$, which is defined as~\cite{Adesso2005,Vitali2007,Clerk2013}:
\begin{equation}
\begin{split}
E_n=\max[0,-\ln(2\zeta)],
\end{split}
\label{eq:negativity}
\end{equation}
where $\zeta$ is the smallest symplectic eigenvalue of the partially transposed covariance matrix $\tilde{\nu}$. It can be obtained from the reduced covariance matrix related to the two resonators $1$ and $2$, i.e., eliminating the first two rows and columns of the full covariance matrix $C$, and replacing $p_j$ with $-p_j$. The symplectic eigenvalue is calculated as the square root of the ordinary eigenvalues of $-(\sigma\tilde{\nu})^2$, where $\sigma=J\oplus J$ and $J$ is a $2\times 2$ matrix with $J_{1,2}=-J_{2,1}=1$ and $J_{1,1}=J_{2,2}=0$.

The numerical results for the covariance matrix are obtained  
by mapping the quantum Heisenberg-Langevin equations into the $c$-number stochastic Langevin equations, whose $k$th trajectory satisfies the set of equations
\begin{widetext}
\begin{equation}
\begin{split}
&^{k}\dot{q}_j=\omega_j{^{k}p_j},\\
&^{k}\dot{p}_j=-\omega_j {^{k}q_j}-\gamma_j{^{k}p_j}+\mathcal{L}_j(Q_1,Q_2,{^{k}q}_1,{^{k}q}_2)\left(\vert {^{k}a}\vert^2-\dfrac{1}{2}\right)+\xi_j,\\
&^{k}\dot{a}=\left\{-\kappa+i\Delta-i\left[\delta\omega\left(Q_1,Q_2,{^{k}q}_1,{^{k}q}_2\right)\right]\right\}{^{k}a}+E_1e^{-i\omega_1t}+E_2e^{i\omega_2t}+\sqrt{2\kappa}a^{in},
\end{split}
\label{eq:quantum langevin en WA}
\end{equation}
\end{widetext}
where the $c$-number noise terms possess the following autocorrelation functions: $\langle a^{in}(t)^*a^{in}(t')\rangle= (\bar{n}_a+{1}/{2})\delta(t-t')$ and $\langle \xi_j(t)\xi_j'(t')\rangle = 2\gamma_j (\bar{n}_j+{1}/{2})\delta_{jj'}\delta(t-t')$, where $n_a =[\exp(\hbar \omega_c/k_B T)-1]^{-1}$, and $n_j =[\exp(\hbar \omega_j/k_B T)-1]^{-1}$ are the corresponding thermal occupancies at temperature $T$.
The covariance matrix elements are obtained by averaging over the $N$ trajectories, $\langle \hat{o}\rangle=\sum_k \frac{1}{N}{^{k}o}$, and 
$\langle \{\hat{o}_j,\hat{o}'_{j'}\}/2\rangle=\sum_k \frac{1}{N} {^{k}o_j} {^{k}o'}_{j'}$, where $\hat{o}$ represents any element of the vector $\hat{u}$.

\begin{figure}[]
\centering
\includegraphics[width=3.2in]{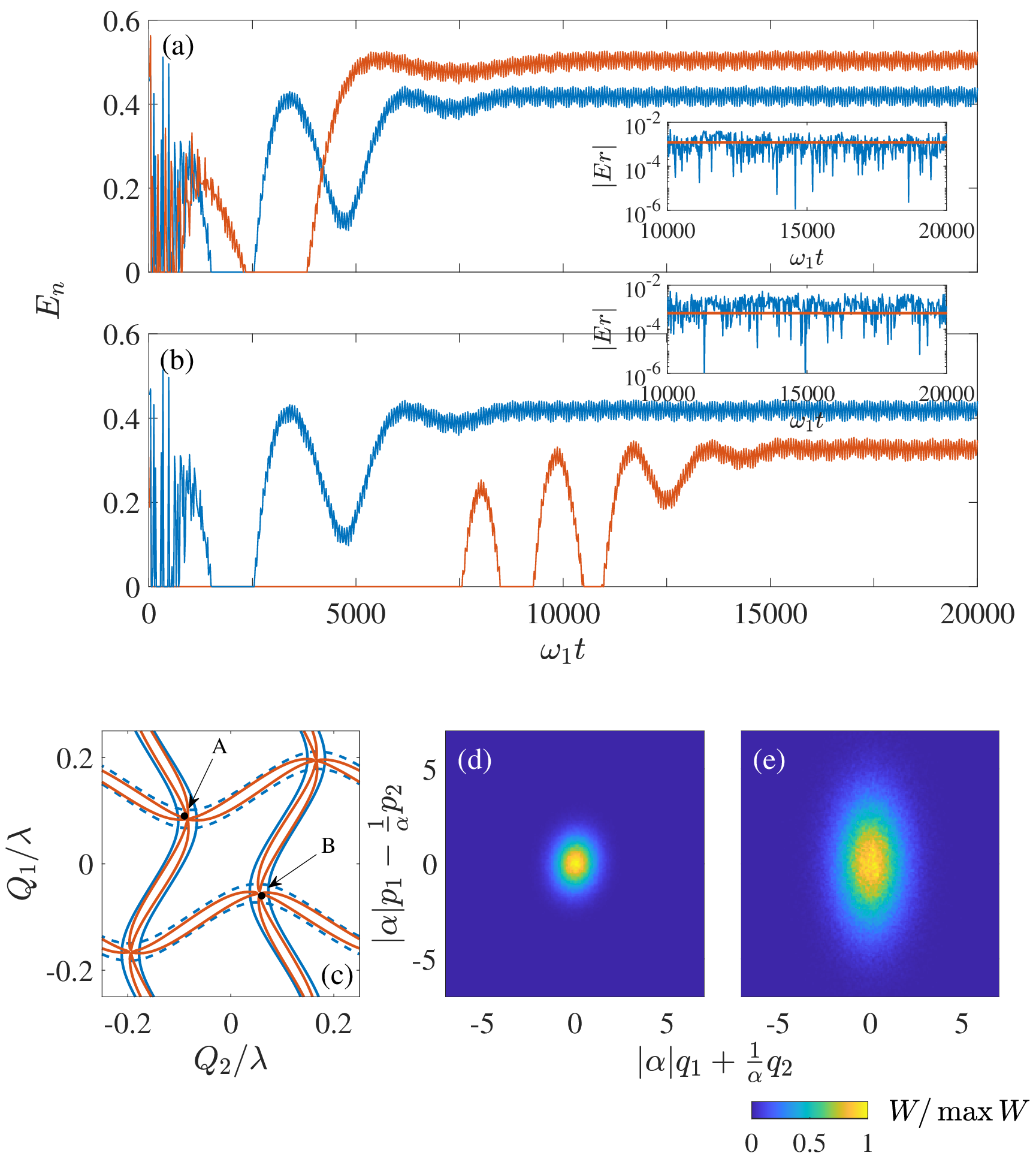}  
\caption{(a), (b) Comparison of the time evolution of the logarithmic negativity $E_n$. The red lines and the blue lines are obtained by $384000$ realizations of the stochastic full dynamics equation Eq.~\eqref{eq:quantum langevin en WA} and the stochastic first-order equation Eq.~\eqref{eq:c langevin fr}, respectively. For the first-order approximated dynamics, the error between the statistical results and the results obtained by the mean field linearized treatment is quantified by the blue lines in the inset of (a) and (b), and the red lines in the insets are the corresponding time-average. (c) Representation of the membranes' positions `A', corresponding to $Q_1/\lambda=-0.09$, $Q_2=-Q_1$, and `B', which is $Q_1/\lambda=0.06$, $Q_2=-Q_1$, on the contour plot of Fig.~\ref{fig:3}(b). (d) and (e) Phase space distribution in the EPR-like variables $\{\vert \alpha \vert q_1+\frac{1}{\alpha}q_2,\{\vert \alpha \vert p_1-\frac{1}{\alpha}p_2\}$, which corresponds to the resonator states in (a) at $\omega_1 t=20000$. The results are obtained using the full dynamics set of equations; we have chosen $\alpha=1$ in (d) and $\alpha=-1$ in (e), corresponding to the squeezed and anti-squeezed pairs of variables of the entangled state. Here we use $\omega_1=235$\,kHz, $\omega_2=1.1\omega_1$, $L=0.009$\,m and $\bar{n}_{a,b_1,b_2}=0$. $E_1$ and $E_2$ is adjusted to ensure $E_1g_1/\omega_1[i(\Delta-\omega_1)+\kappa]=12163.6$ and  $E_2g_2/\omega_1[i(\Delta+\omega_2)+\kappa]=45180.3$ always hold. The initial states of the three modes are set as vacuum states and the other parameters are the same as those in Fig.~\ref{fig:2}.
\label{fig:6}}
\end{figure}

In Fig.~\ref{fig:6}(a) we plot the time evolution of the logarithmic negativity of the two mechanical resonators $E_n$, which is obtained by $384000$ realizations of the full dynamics equation~\eqref{eq:quantum langevin en WA}. In order to enhance the entanglement, we chose a different set of parameters with respect to the previous section, corresponding to a shorter optical cavity ($L=0.009$\,m), zero temperature, ($\bar{n}_{a}=\bar{n}_{1}=\bar{n}_{2}=0$), and a larger frequency difference between the resonators ($\omega_2=1.1\omega_1$), which is a condition which is required for the reservoir engineering scheme to work. To provide a comparison, Fig.~\ref{fig:6}(a) also shows the evolution of the entanglement measure, obtained by calculating the stochastic Langevin equations with only first-order optomechanical couplings and the same number of simulations (see Appendix \ref{Mean-field treatment of the first-order quantum Langevin equation} for the details of the Langevin equations and the later mentioned mean-field linearized theory). The parameter regime under consideration is in the region where higher-order couplings are no longer negligible [point `A' in Fig.~\ref{fig:6}(c)], and we see that the degree of entanglement exceeds the prediction of the first-order coupling equations, specifically with an increment $\Delta E_n\simeq 0.06$. Note that the displacements of the resonators remain negligible throughout the entire process, differently from what happens in Sec. IV. This fact indicates that the entanglement enhancement originates from the long-range direct interaction term between the two membranes, whose effective strength $g_t\langle\hat{a}^\dagger \hat{a}\rangle$ is proportional to the number of intra-cavity photons. The first-order quantum Langevin equations can also be treated by mean-field theory in order to obtain the corresponding covariance matrix. Hence, the discrepancy between the results obtained by these two methods, defined as $Er$, is employed to assess the accuracy of the stochastic Langevin equations. As illustrated in the inset of Fig.~\ref{fig:6}(a), for a number of realizations $N=384000$, the average standard deviation of the stochastic Langevin equation for calculating $E_n$ is $\sigma(Er)\simeq 0.0012$, which is only $1/50$ of the entanglement increment mentioned above. Following the far exceeded $5\sigma$ criterion, we confirm that the entanglement enhancement predicted by the full dynamics is not a statistical fluctuation. In ~\ref{fig:6}(b), we present the inverse scenario, wherein higher-order couplings diminish the entanglement between resonators, when they are placed at the position corresponding to point `B'. Furthermore, the data obtained by Eq.~\eqref{eq:quantum langevin en WA} enable us to reconstruct the distribution function of the entangled resonators in the phase space, as shown in Figs.~\ref{fig:6}(d) and (e). In this case, the EPR-like variables $\{q_1+q_2,p_1-p_2\}$, $\{q_1-q_2,p_1+p_2\}$ are selected as the basis vectors of the phase space instead of $\{q_1,p_1\}$ and $\{q_2,p_2\}$. As illustrated in Fig.~\ref{fig:6}(d), the fluctuation distribution is compressed to a value $\langle[\Delta (\hat{q}_1+\hat{q}_2)]^2\rangle+\langle[\Delta (\hat{p}_1-\hat{p}_2)]^2\rangle <2$, which violates the Duan separability criterion~\cite{Duan2000}. Correspondingly, the fluctuation is amplified along the $q_1-q_2$ and $p_1+p_2$ directions. In summary, this example illustrates that working with the full dynamics equations is useful for performing a more accurate analysis of the OMS system even for what concerns the quantum properties, including mechanical Gaussian entanglement.

\section{Discussion of other situations}
\label{Discussion of other situations}
\begin{figure}[]
\centering
\includegraphics[width=3.2in]{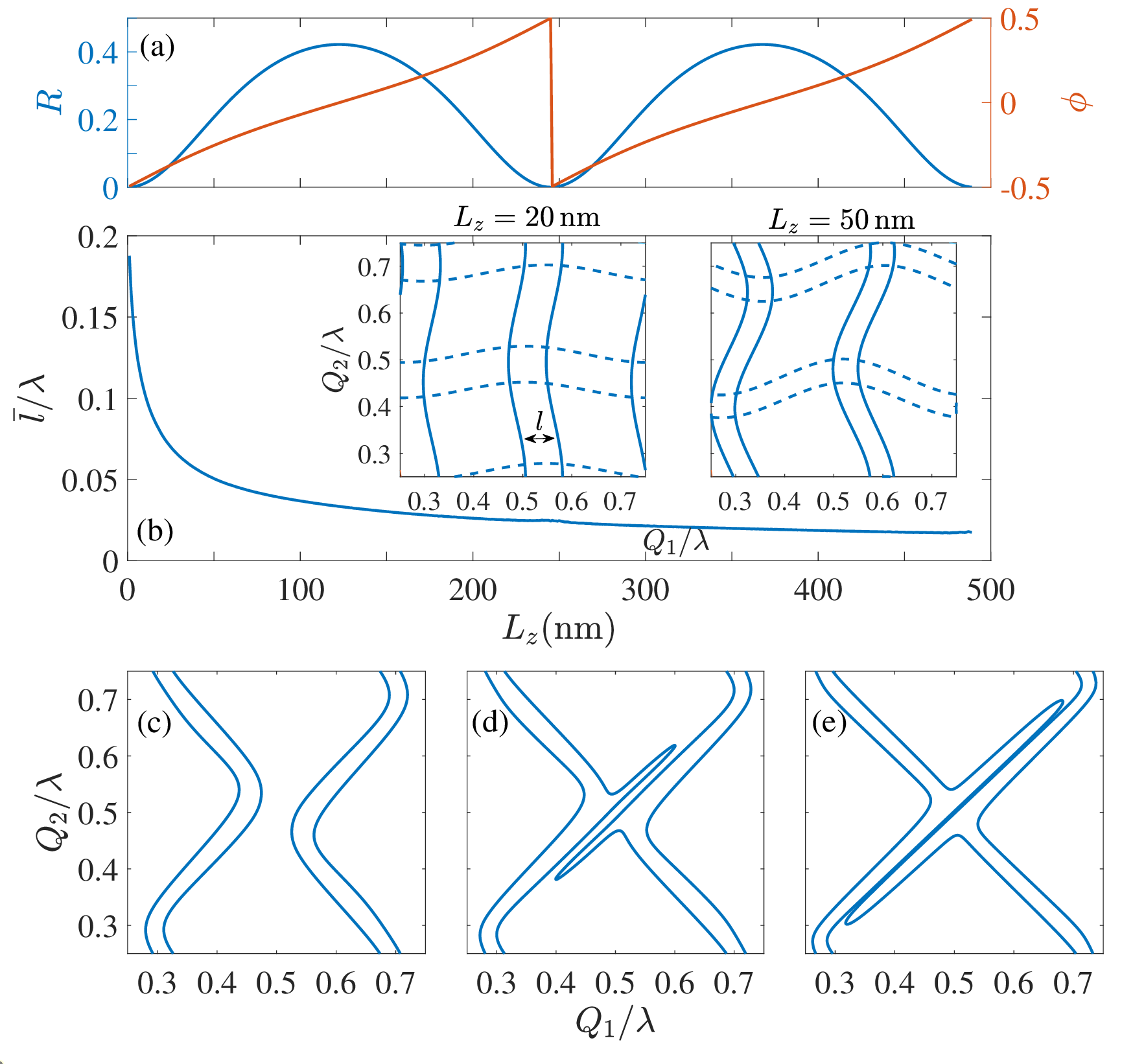}  
\caption{(a) Modulus and phase of the complex reflectivity $r=\sqrt{R}e^{i\phi}$ as a function of the thickness of the membrane $L_z$. (b) The average width of the $\#$-shaped region where high-order coupling cannot be ignored, characterized by the condition $\vert g_{j,2}/g_{j}\vert>10^{-7.5}$, as a function of the thickness of the membrane $L_z$. The two insets illustrate this region for $L_z=20$\,nm on the left, and $L_z=50$\,nm on the right, respectively. (c), (d),(e) Contour lines of $\vert g_{1,2}/g_{1}\vert=10^{-7.5}$ for $R=0.7$, $R=0.8$ and $R=0.9$, respectively. In (c), (d) and (e), we set $\phi=0$ and $L_z=104$\,nm since Eq.~\eqref{eq:R} becomes ineffective. The other parameters are the same as those in Fig.~\ref{fig:2}.
\label{fig:7}}
\end{figure}

In the preceding section, we identified the region in which the higher-order coupling cannot be disregarded in the position plane $Q_1$-$Q_2$, which assumes a $\#$-shape. Referring to Eq.~\eqref{eq:R}, it can be determined that the extent of these regions depends upon the membrane reflectivity $r=\sqrt{R}e^\phi$ and upon the mechanical zero-point fluctuation amplitude $x_{\text{ZPF},j}$. The variables $R$ and $\phi$ vary periodically with the membrane thickness $L_z$, as illustrated in Fig~\ref{fig:7}(a). Conversely, $x_{\text{ZPF},j}$ decreases monotonically with $L_z$, since thicker membranes are associated with larger masses. The inset in Fig.~\ref{fig:7}(b) presents the higher-order significant region for membrane thicknesses $L_z=20$\,nm and $50$\,nm, respectively. The extent of this region is larger than the one shown in Fig.~\ref{fig:3}(b), especially for the thinner membrane ($L_z=20$\,nm). We calculate the (position) average of the width of the region, $\bar{l}$, to characterize the extension of the region. We describe quantitatively the dependence of the extent of the $\#$-shape region on the membrane thickness, as shown in Fig.~\ref{fig:7}(b), where a decreasing monotonic trend is pointed out. The strongest variation is found in the range $L_z<50$\,nm. Furthermore, although the reflectivity of a standard homogeneous membrane usually has values $R<0.5$, as illustrated in Fig.~\ref{fig:7}(a), it can even reach $R>0.99$ by adopting patterned sub-wavelength grating or photonic-crystal membranes~\cite{Stambaugh2015,Norte2016}, where Eq.~\eqref{eq:R} does not apply. In this scenario, we keep the membrane thickness to $L_z = 104$\,nm, as in the previous discussion, and we adjust the reflectivity to $R=0.7$,~$0.8$, and~$0.9$. In Figs.~\ref{fig:7}(c), (d) and (e), we plot the contour lines corresponding to $\vert g_{1,2}/g_1\vert=10^{-7.5}$, which is the same criterion for establishing the region of the vertical strokes in the $\#$-shaped area in Fig.~\ref{fig:3}. The contour lines for $R\ge 0.8$ exhibit different curvature and encompass a different domain with respect to the case of low reflectivity membrane. We note that the influence of second-order coupling is more pronounced in this instances.

We point out that the aim of our derivation is to obtain the full dynamics equations from the renormalized cavity field, whose universality extends beyond two-body systems. Building upon the conclusions of Xuereb \textit{et. al.}~\cite{Xuereb2012}, who have already obtained the renormalized frequency of multiple oscillators to the cavity field as a function of the oscillator ensemble, i.e., $\delta\omega(x_1,x_2,...)$, one could obtain the multi-body OMS dynamics equations for all interactions, including higher-order coupling and long-range interactions. Another interesting case is the dissipative OMS of Ref.~\cite{Xuereb2011}, whose higher-order coupling has not been discussed in the literature, and that could be afforded with the same method described here. 

\section{Conclusion}
\label{Conclusion}
In summary, we have studied the full dynamics equations for the ``sandwich-in-the-middle'' OMS, obtained from the exact solution of the cavity mode frequency shift as a function of the position of the two membranes, and including all orders in the membrane displacements. We have seen that it is possible to find system configurations in which the usual treatment in which the optomechanical interaction is described at first order in the membrane displacement does not provide an accurate description of the mechanical resonator dynamics. We have shown two different examples of the inadequacy of the traditional approach. First, we focused on the scenario where the OMS is driven by a blue detuning, thereby inducing the membrane resonators to reach a state of self-sustained oscillation. We show that, thanks to the large amplitude of the limit cycle oscillations, there are system configurations, i.e., membrane positions along the cavity axis, where the synchronization dynamics given by the full equations including all orders of the resonator displacements significantly deviates from the predictions of the standard OMS equations, which stop at first order. 

In a second example we considered the quantum dynamics associated with the generation of Gaussian entangled states of the two resonators, by means of a reservoir engineering scheme based on a suitable bichromatic driving of the cavity mode. 
Also in this case we have found parameter regimes in which one has an appreciable difference between the predictions of the exact full dispersive  optomechanical interaction and those based on the traditional first-order treatment of the interaction. 

The present work provides a more accurate description of OMS, thereby enabling exact predictions for the dynamical properties of OMS in a wider range of parameters. In addition, it can provide a basis for the application of OMS in quantum information processing, quantum precision measurement, and other fields. We expect that under the full dynamics description of the sandwich-in-the-middle setup, other physical phenomena, not discussed in this paper, could be affected, such as the nonlinear dynamics of a two-membrane optomechanical setup~\cite{Piergentili2021}, or the dark mode effect on two mechanical modes~\cite{Huang2023}. The full dynamics of OMS in a structured environment is a topic which deserves investigation and will be the subject of further research.
\begin{acknowledgements}
W.~L. is supported by the National Natural Science Foundation of China (Grants No.~12304389), the Scientific Research Foundation of NEU (Grant No. 01270021920501*115). 
F.~M., P.~P., F.~R., D.~V. acknowledge financial support from NQSTI within PNRR MUR Project PE0000023-NQSTI.
\end{acknowledgements}
\appendix
\begin{widetext}
\section{The renormalized dissipation rate is in the common model}
\label{The renormalized dissipation rate is in the common model}
The coefficients of Eqs.~\eqref{eq:L1} and~\eqref{eq:L2} are expressed as follows:
\begin{equation}
\begin{split}
&f_1=(2-R+2R^2) \cos(2k\tilde{q_1}-\phi),\\
&f_2=R(\cos[2k (\tilde{q_1} - 2 \tilde{q_2})-3\phi] - 3 \cos(2k\tilde{q_2}+\phi) -\cos[2k(\tilde{q_2}-2\tilde{q_1})+3\phi]),\\
&f_3=(1+R^2-2R\cos[2 k(\tilde{q_2}-\tilde{q_1})+2\phi])^{3/2},\\
&f_4=\sqrt{1-\dfrac{4 R \cos^2[k (\tilde{q_1}+\tilde{q_2})] \sin^2[k(\tilde{q_2}-\tilde{q_1})+\phi]}{1+R^2-2R\cos[2 k(\tilde{q_2}-\tilde{q_1})+2\phi]}},\\
&f_5=-2(1+R^2) \cos[2 k (\tilde{q_2}-\tilde{q_1})+2\phi]+R(3+\cos[4 k(\tilde{q_2}-\tilde{q_1})+4\phi]),\\
&f_6=\sqrt{\dfrac{(-1 + R \cos[2 k(\tilde{q_2}-\tilde{q_1})+2\phi])^2}{1+R^2-2R\cos[2 k(\tilde{q_2}-\tilde{q_1})+2\phi]}},\\
&f_7=(-2+R-2R^2) \cos(2k\tilde{q_2}+\phi),\\
&f_8=R(\cos[2k (\tilde{q_1} - 2 \tilde{q_2})-3\phi] + 3 \cos(2k\tilde{q_1}-\phi) -\cos[2k(\tilde{q_2}-2\tilde{q_1})+3\phi]),
\end{split}
\label{eq:fs}
\end{equation}
\end{widetext}

\section{Mean-field treatment of the first-order quantum Langevin equation}
\label{Mean-field treatment of the first-order quantum Langevin equation}
The quantum Langevin equation  corresponding to the reservoir-engineering scheme in the main text with only the radiation pressure interaction is:
\begin{equation}
\begin{split}
&\dot{\hat{q}}_j=\omega_j{\hat{p}_j},\\
&\dot{\hat{p}}_j=-\omega_j {\hat{q}_j}-\gamma_j{\hat{p}_j}+g_j\hat{a}^\dagger\hat{a}+\hat{\xi}_j,\\
&\dot{\hat{a}}=\left(-\kappa+i\Delta'\right)\hat{a}+ig_1\hat{a}\hat{q}_1+ig_2\hat{a}\hat{q}_2\\&
\,\,\,\,\,\,\,+E_1e^{-i\omega_1t}+E_2e^{i\omega_2t}+\sqrt{2\kappa}\hat{a}^{in},
\end{split}
\label{eq:quantum langevin fr}
\end{equation}
and the corresponding stochastic Langevin equation is:
\begin{equation}
\begin{split}
&^{k}\dot{q}_j=\omega_j{^{k}p_j},\\
&^{k}\dot{p}_j=-\omega_j {^{k}q_j}-\gamma_j{^{k}p_j}+g_j\left(\vert {^{k}a}\vert^2-\dfrac{1}{2}\right)+\xi_j,\\
&^{k}\dot{a}=\left(-\kappa+i\Delta'\right)a+ig_1{^{k}a}{^{k}q_1}+ig_2{^{k}a}{^{k}q_2}\\&
\,\,\,\,\,\,\,+E_1e^{-i\omega_1t}+E_2e^{i\omega_2t}+\sqrt{2\kappa}a^{in}.
\end{split}
\label{eq:c langevin fr}
\end{equation}
As mentioned in the main text, the mean field method based on linearized approximation can also be used to analyze this equation. Specifically, each operator is written as the sum of its expectation value and fluctuation, that is, $\hat{o}=\langle \hat{o}\rangle +\delta o$~\cite{Vitali2007,Wang2014}. Here, $o$ represents any element in $\hat{u}$ as defined in the main text. Substituting this expression into Eq.~\eqref{eq:quantum langevin fr} yields two equations, one that is satisfied by the expectation value of the operator:
\begin{equation}
\begin{split}
&\langle\dot{\hat{q}}\rangle_j=\omega_j\langle \hat{p}\rangle_j,\\
&\langle\dot{\hat{p}}\rangle_j=-\omega_j \langle \hat{q}\rangle_j-\gamma_j{\langle \hat{p}\rangle_j}+g_j\vert \langle \hat{a}\rangle\vert^2,\\
&\langle\dot{a}\rangle=\left(-\kappa+i\Delta'\right)\langle \hat{a}\rangle+ig_1\langle \hat{a} \rangle\langle \hat{q}\rangle_1+ig_2\langle \hat{a} \rangle\langle \hat{q}\rangle_2\\&
\,\,\,\,\,\,\,\,+E_1e^{-i\omega_1t}+E_2e^{i\omega_2t}+\sqrt{2\kappa}\hat{a}^{in},
\end{split}
\label{eq:mean langevin fr}
\end{equation}
 and another that is satisfied by its fluctuation:
\begin{equation}
\begin{split}
&\dot{\delta{q}}_j=\omega_j{\delta p_j},\\
&\dot{\delta{p}}_j=-\omega_j \delta{q}_j-\gamma_j{\delta p_j}+g_j\left(\langle \hat{a}\rangle\delta{a}^\dagger+\langle \hat{a}\rangle^*\delta{a}\right)+\hat{\xi}_j,\\
&\dot{\delta a}=\left(-\kappa+i\Delta'\right)\delta{a}+ig_1\left(\langle \hat{a}\rangle \delta{q}_1+\langle \hat{q}\rangle_1\delta{a}\right)\\
&\,\,\,\,\,\,\,+ig_2\left(\langle \hat{a}\rangle \delta{q}_2+\langle \hat{q}\rangle_2\delta{a}\right)+\sqrt{2\kappa}\hat{a}^{in}.
\end{split}
\label{eq:f langevin fr}
\end{equation}
Here the correlation terms are approximated as: $\langle \hat{o}\hat{o}'\rangle\simeq\langle \hat{o}\rangle\langle\hat{o}'\rangle$, and all high-order fluctuation terms are ignored. The Eq.~\eqref{eq:f langevin fr} is written in a more compact form as $\dot{\delta{u}}=S\delta{u}+\zeta$, where the vectors are set as $\delta\hat{u}=(\delta x,\delta y, \delta q_1,\delta p_1, \delta q_2,\delta p_2)^{\top}$ and $\zeta=(x^{in},y^{in}, 0,\xi_1, 0,\xi_2)^{\top}$, with $x^{in}=(\hat{a}^{\dagger,in}+\hat{a}^{in})/\sqrt{2}$ and  $y^{in}=(\hat{a}^{\dagger,in}-\hat{a}^{in})/i\sqrt{2}$. Then one can derive the following equation satisfied by the covariance matrix~\cite{Wang2014}:
\begin{equation}
\begin{split}
\dfrac{d}{dt}C=SC+CS^{\top}+N,
\end{split}
\label{eq:C langevin fr}
\end{equation}
where $N=\text{diag}[-\kappa(2\bar{n}_{a}+1),-\kappa(2\bar{n}_{a}+1),0,\gamma_1(2\bar{n}_{b,1}+1),0,\gamma_2(2\bar{n}_{b,2}+1)]$ is the diagonal matrix  corresponding to the noise operator, and $S$ is the time-dependent coefficient matrix: 
\begin{equation}
\begin{split}
S=\begin{pmatrix}
 S_{o} & S_{om_1} & S_{om_2} \\
 S_{m_1o} & S_{m_1} & S_{m_1m_2} \\
 S_{m_2o} & S_{m_2m_1} & S_{m_2} \\
\end{pmatrix},
\end{split}
\label{eq:SMatrix}
\end{equation}
with 
\begin{equation}
\begin{split}
&S_{o}=\begin{pmatrix}
 -\kappa & \Delta''  \\
 -\Delta''& -\kappa  \\
\end{pmatrix},
S_{m_j}=\begin{pmatrix}
 0 & \omega_j  \\
 -\omega_j& -\gamma_j  \\
\end{pmatrix},
\\
&S_{om_j}=\begin{pmatrix}
 -\sqrt{2}g_j\text{Im}(\langle \hat{a}\rangle) & 0  \\
 \sqrt{2}g_j\text{Re}(\langle \hat{a}\rangle)& 0  \\
\end{pmatrix},\\
&S_{m_jo}=\begin{pmatrix}
 0 & 0  \\
 \sqrt{2}g_j\text{Re}(\langle \hat{a}\rangle)& \sqrt{2}g_j\text{Im}(\langle \hat{a}\rangle)
\end{pmatrix},\\
\end{split}
\label{eq:SMatrix}
\end{equation}
and $S_{m_1m_2}=S_{m_2m_1}=0$. Here $\Delta''=\Delta'-g_1q_1-g_2q_2$. The logarithmic negativity obtained by statistical analysis of the result of Eq.~\eqref{eq:c langevin fr} is denoted as $E_{n_s}$, and that obtained according to the mean-field Eqs.~\eqref{eq:mean langevin fr} and~\eqref{eq:f langevin fr} is denoted as $E_{n_m}$, then the error $Er$ in the main text is defined as:
\begin{equation}
\begin{split}
Er(t)=E_{n_s}(t)-E_{n_m}(t).
\end{split}
\label{eq:Er}
\end{equation}
Its time average is approximately zero, thus its standard deviation in the time interval $[t_1, t_2]$ can be calculated:
\begin{equation}
\begin{split}
\sigma(Er)=\dfrac{1}{t_2-t_1}\int_{t_1}^{t_2}\vert Er(t)\vert dt,
\end{split}
\label{eq:Er}
\end{equation}
which is adopted to evaluate the algorithm's accuracy.


\begin{thebibliography}{35}
\bibitem{exp2} In fact, the input term should be comprised of three distinct components: $\sqrt{2\kappa_\text{in}}a^{in,1}+\sqrt{2\kappa_\text{in}}a^{in,2}+\sqrt{2\kappa_\text{ex}}a^{in,ex}$. We are replacing the sum of optical noises as $\sqrt{2\kappa}a^{in}$ due to the negligible difference between these two noise terms corresponding to vacuum noise.
\bibitem{exp1} It can be checked by the antisymmetric structure $\mathcal{L}_1(\tilde{q}_1\rightarrow\tilde{q}_2)=-\mathcal{L}_2$ when $\phi=0$.
\end{thebibliography}
\end{document}